\documentclass[11pt,a4paper]{amsart}

\newcommand {\Cee}    {{\mathbb  C}}
\newcommand {\Zee}    {{\mathbb  Z}}

\newcommand {\fcvect}   {{\mathfrak{cvect}}}
\newcommand {\fg}     {{\mathfrak{g}}}    %
\newcommand {\fle}    {{\mathfrak{le}}}
\newcommand {\fsle}   {{\mathfrak{sle}}}
\newcommand {\fsvect} {{\mathfrak{svect}}}
\newcommand {\fu}     {{\mathfrak{u}}}
\newcommand {\fvect}  {{\mathfrak{vect}}}   %
\newcommand {\fk}     {{\mathfrak{k}}}
\newcommand {\fh}     {{\mathfrak{h}}}
\newcommand {\fsl}     {{\mathfrak{sl}}}
\newcommand {\fgl}     {{\mathfrak{gl}}}
\newcommand {\fder}   {{\mathfrak{der}}}
 \newcommand {\fspe}     {{\mathfrak{spe}}}

\newcommand{\rmname}[1]
  {\expandafter\newcommand \csname #1\endcsname {{\operatorname{#1}}}}

\newcommand{\rmnameii}[2]
  {\expandafter\newcommand \csname #1\endcsname {{\operatorname{#2}}}}

\rmnameii{Div}{div}
\rmname{Dim}
\rmname{Hom}
\rmname{Le}
\rmname{ad}
\rmname{im}
\rmname{sign}

\newcommand {\ev} {{\bar0}}
\newcommand {\od} {{\bar1}}
\newcommand {\degree}  {{}^\circ}
\newcommand {\pder}[1] {{\frac{\partial}{\partial {#1}}}}
\newcommand {\pderf}[2] {{\frac{\partial {#1}}{\partial {#2}}}}

\newcommand {\secno} {}
\newcommand {\ssecfont} {\normalfont\bf}

\newtheorem{Theorem}{\secno Theorem$\!\!$}

\newtheorem{Lemma}[Theorem]{\secno Lemma$\!\!$}

\theoremstyle{definition}

\newcommand {\ssec}{\subsection*}

\newcommand {\ssbegin}[2]
  {\def \secno {\gdef \secno {}{\ssecfont #1. }}%
   \begin{#2}}

\newenvironment {rem*}[1]
    {\gdef\thname{#1{\!\!}} \begin{remn}}%
    {\end{remn}}
\newtheorem{remn}[Theorem]{\thname}

\begin{document}

\title[Explicit bracket in an exceptional Lie superalgebra]
 {Explicit bracket in the exceptional simple Lie superalgebra
  {\Large $\fcvect(0|3)_*$}}

\author{Irina Shchepochkina \and Gerhard Post}

\address{I.Shch. : on leave of absence from the Independent University of
Moscow.
Correspondence: c/o D.~Leites, Department of Mathematics, University
of Stockholm, Roslagsv. 101, Kr\"aftriket hus 6, S-106 91, Stockholm, Sweden.}
\address{G.P. : Department of Applied Mathematics, University Twente,\\
  P.O. Box 217, 7500 AE Enschede, The Netherlands.}

%%\keywords {Lie superalgebra, Cartan prolongation.}
%%\subjclass{17A70}

\begin{abstract}  This note is devoted to a more detailed
  description of one of the five simple exceptional Lie superalgebras
  of vector fields, $\fcvect(0|3)_*$, a subalgebra of $\fvect(4|3)$.
  We derive differential equations for its elements, and solve these
  equations. Hence we get an exact form for the elements of
  $\fcvect(0|3)_*$. Moreover we realize $\fcvect(0|3)_*$ by "glued"
  pairs of generating functions on a $(3|3)$-dimensional periplectic
  (odd symplectic) supermanifold and describe the bracket
  explicitly.
  \\[1mm]
\end{abstract}

\thanks{I.Shch. expresses her thanks: to D.~Leites for rising the
  problem and help; to RFBR grant 95-01-01187 and NFR (Sweden) for
  part of financial support; University of Twente and Stockholm
  University for hospitality; to P.~Grozman whose computer experiments
  encouraged her to carry on with unbearable calculations.}

\maketitle

\begin{center}
March 3, 1997
\end{center}

\section*{Introduction}

V. Kac \cite{K} classified simple finite-dimensional Lie superalgebras
over ${\Cee}$. Kac further conjectured \cite{K} that passing to
infinite-dimensional simple Lie superalgebras of vector fields with
polynomial coefficients we only acquire the straightforward analogues
of the four well-known Cartan series: $\fvect(n)$, $\fsvect(n)$,
$\fh(2n)$ and $\fk(2n+1)$ (of all, divergence-free, Hamiltonian and
contact vector fields, respectively, realized on the space of
dimension indicated).

It soon became clear \cite{L1}, \cite{ALSh}, \cite{L2}, \cite{L3} that
the actual list of simple vectoral Lie superalgebras is much larger.
Several new series were found.

Next, exceptional vectoral algebras were discovered \cite{Sh1}, \cite{Sh2}; for
their detailed description see \cite{Sh3}, \cite{GLSh}. All of them are
obtained
with the help of a Cartan prolongation or a generalized prolongation,
cf. \cite{Sh1}. This description is, however, not always satisfactory; a
more succinct presentation (similar to the one via generating functions
for the elements of $\fh$ and $\fk$) and a more explicit formula for
their brackets is desirable.

The purpose of this note is to give a more lucid description of one of
these exceptions, $\fcvect(0|3)_*$. In particular we offer a
multiplication table for $\fcvect(0|3)_*$ that is simpler than
previous descriptions, by use of "glued" pairs of generating functions
for the elements of $\fcvect(0|3)_*$.

This note can be seen as a supplement to \cite{Sh3}. To be self-contained
and to fix notations we introduce some basic notions in section
0.

Throughout, the ground field is ${\Cee}$.

\section*{\protect\S 0. Background}

\ssec{0.1} We recall that a superspace $V$ is a $\Zee /2$-graded
space; $V=V_{\ev}\oplus V_{\od }$. The elements of $V_{\ev}$ are
called {\it even}, those of $V_{\od }$ {\it odd}. When considering an
element $x\in V$, we will always assume that $x$ is homogeneous,
i.e. $x\in V_{\ev}$ or $x\in V_{\od}$. We write $p(x)={\bar{i}}$ if
$x\in V_{\bar{i}}$. The superdimension of $V$ is $(n|m)$, where
$n=\dim(V_{\ev}$) and $m=\dim(V_{\od}$).

For a superspace $V$, we denote by $\Pi (V)$ the same
superspace with the shifted parity, i.e., $\Pi(V_{\bar i})= V_{\bar
  i+\od}$.

\ssec{0.2} Let $x=(u_1,\dots,u_n,\xi_1,\dots,\xi_m)$, where
$u_1,\dots,u_n$ are even indeterminates and $\xi_1,\dots,\xi_m$ odd
indeterminates. In the associative algebra $\Cee[x]$ we have that
$x\cdot y = (-1)^{p(x)p(y)}y\cdot x$ (by definition) and hence
$\xi_i^2=0$ for all $i$. The derivations $\fder(\Cee[x])$ of $\Cee[x]$
form a Lie superalgebra; its elements are vector fields. These
polynomial vector fields are denoted by $\fvect(n|m)$. Its elements
are represented as
\[
D=\sum\limits_i f_i\pder{u_{i}} + \sum\limits_j g_j\pder{\xi_{j}}
\]
where $f_i \in \Cee[x]$ and $g_j \in \Cee[x]$ for all $i,j=1..n$.
We have $p(D)=p(f_i)=p(g_j)+\od$ and the Lie product is given by the
commutator
\[ [D_1,D_2] = D_1 D_2 - (-1)^{p(D_1)p(D_2)} D_2 D_1. \]
On the vector fields we have a map, $\Div : \fvect(n|m) \to \Cee[x]$,
defined by
\[ \Div D= \Div(\sum\limits_{i=1}^n f_i\pder{u_{i}} +
                \sum\limits_{j=1}^n g\pder{\xi_{j}})
             =  \sum\limits_{i=1}^n \pderf{f_{i}}{u_{i}} -
    (-1)^{p(D)} \sum\limits_{j=1}^n \pderf{g_{j}}{\xi_{j}}.
\]
A vector field $D$ that satisfies $\Div D = 0$ is called {\it special}.
The linear space of special vector fields in $\fvect(n|m)$ forms a Lie
superalgebra, denoted by $\fsvect(n|m)$.

\ssec{0.3} Next we discuss the Lie superalgebra of Leitesian vector
fields $\fle(n)$. It consists of the elements $D \in \fvect(n|n)$ that
annihilate the 2-form $\omega=\sum_i d u_i d \xi_i$. Hence $\fle(n)$
is an odd superanalogon of the Hamiltonian vector fields (in which
case $\omega=\sum_i d p_i d q_i$). Similar to the Hamiltonian case,
there is a map $\Le : \Cee[x] \to \fle(n)$, with
$x=(u_1,\dots,u_n,\xi_1,\dots,\xi_n)$:
\[ \Le_f=\sum\limits_{i=1}^n( \pderf{f}{u_i}\ \pder{\xi_i} +(-1)^{p(f)}
                           \pderf{f}{\xi_i}\ \pder{u_i}) \]
Note that $\Le$ maps odd elements of $\Cee[x]$ to even elements of
$\fle(n)$ and vice versa. Moreover Ker$(\Le) = \Cee$.
We turn $\Cee[x]$ (with shifted parity) into a Lie superalgebra with
(Buttin) bracket $\{f,g\}$ defined by
\[ \Le_{\{f,g\}} = [\Le_f,\Le_g] \]
A straightforward calculation shows that
\[ \{ f, g\} =\sum\limits_{i=1}^n \ (\pderf{f}{u_i}\ \pderf{g}{\xi_i}
              +(-1)^{p(f)}\ \pderf{f}{\xi_i}\ \pderf{g}{u_i}). \]
This way $\Pi \Cee[x]/\Cee\cdot 1$ is a Lie superalgebra isomorphic to
$\fle(n)$. We call $f$ the generating function of $\Le_f$. {\it Here and
throughout} $p(f)$ will denote the parity in $\Cee[x]$, not in $\Pi
\Cee[x]$. So $p(f)$ is the parity of the number of $\xi$ in a
term of $f$.

\ssec{0.4} The algebra $\fle(n)$ contains certain important
subalgebras. First of all there is $\fsle(n)$, the space of special Leitesian
vector fields:
\[ \fsle(n) = \fle(n) \cap \fsvect(n|n). \]
We have seen that if $D\in \fle(n)$ then $D=\Le_f$ for some $f\in
\Cee[x]$.  Now $D\in \fsle(n)$ iff $f$ is harmonic in the following
sense
\[ \Delta(f):= \sum\limits_{i=1}^n \frac{\partial^2 f}{\partial u_i
  \partial \xi_i} = 0
\]
Usually we simply say $f\in \fsle(n)$, identifying $f$ and $\Le_f$.
This $\Delta$ satisfies the condition $\Delta^2=0$ and hence
$\Delta : \fle(n) \to \fsle(n)$. The image $\Delta(\fle(n)) =:
\fsle\degree(n)$ is an ideal of codimension 1 on $\fsle(n)$.  This
ideal, $\fsle\degree(n)$, can also be defined by the exact sequence
\[ 0\longrightarrow \fsle \degree(n)\longrightarrow \fsle (n)\longrightarrow
\Cee\cdot \Le_{\xi_1\dots\xi_n} \longrightarrow 0. \]
Note that if $\Phi=\sum u_i\xi_i$ and $f\in \fsle(n)$, then
\[ \Delta(\Phi f)= (n +\deg_u f-\deg_\xi f)\cdot f \]
Let $\nu(f)= n +\deg_u f-\deg_\xi f$. Then $\nu(f)\neq 0$ iff $f \in
\fsle\degree(n)$. So on $\fsle\degree(n)$ we can define the right inverse
$\Delta^{-1}$ to $\Delta$ by the formula
\[ \Delta^{-1}f=\frac{1}{\nu(f)}(\Phi f). \]

\ssec{0.5. Cartan prolongs}

We will repeatedly use Cartan prolongation. So let us recall the
definition. Let $\fg$ be a Lie superalgebra and $V$ a $\fg$-module.
Set $\fg_{-1} = V$, $\fg_0 = \fg$ and for $i > 0$ define the $i$-th
{\it Cartan prolong} $\fg_i$ as the space of all $X\in \Hom(\fg_{-1},
\fg_{i-1})$ such that
\[
      X(w_0)(w_1,w_2,\ldots,w_i) = (-1)^{p(w_0)p(w_1)}
      X(w_1)(w_0,w_2,\ldots,w_i)  \]
for all $w_0,\ldots,w_i\in \fg_{-1}$.

The {\it Cartan prolong} (the result of Cartan's {\it prolongation})
of the pair $(V, \fg)$ is $(\fg_{-1}, \fg_{0})_* = \oplus_{i\geq -1}
\fg_i$.

Suppose that the $\fg_0$-module $\fg_{-1}$ is faithful. Then
\[
(\fg_{-1}, \fg_{0})_*\subset \fvect (n|m) = \fder(\Cee[x]),\;
\text{ where }\; n = \dim(V_{\ev}) \text{ and }m = \dim(V_{\od})
\]
and $x=(u_1,\dots,u_n,\xi_1,\dots,\xi_m)$. We have for $i\geq 1$
\[
\fg_i = \{D\in \fvect(n|m): \deg D=i, [D, X]\in\fg_{i-1}\text{ for any }
X\in\fg_{-1}\}.
\]
The Lie superalgebra structure on
$\fvect (n|m)$ induces one on $(\fg_{-1}, \fg_{0})_*$. This way the
commutator of vector fields $[g,v]$, corresponds to the action $g\cdot
v$, $g\in\fg$ and $v\in V$.

We give some examples of Cartan prolongations. Let $\fg_{-1}=V$ be an
$(n|m)$-dimensional superspace and $\fg_0=\fgl(n|m)$ the space of all
endomorphisms of $V$. Then $(\fg_{-1},\fg_0)_*=\fvect(n|m)$. If one
takes for $\fg_0$ only the supertraceless elements $\fsl(n|m)$, then
$(\fg_{-1},\fg_0)_*=\fsvect(n|m)$, the algebra of vector fields with
divergence 0.

\section*{\S 1. The structure of \protect $\fvect (0|3)_*$}

\ssec{1.1} In this note our primary interest is in a certain Cartan
prolongation (denoted by $\fvect(0|3)_*$) and the extension
$\fcvect(0|3)_*$ thereof. Here we will discuss $\fvect(0|3)_*$. Now
$\fvect(0|3)_*$ is a short-hand notation for the Cartan prolongation
with
\[ V=\fg_{-1} = \Pi \Lambda(\eta_1,\eta_2,\eta_3)/\Cee \text{ and }
   \fg_0 =\fder V \]
So $V$ is a superspace of dimension $(4|3)$, with
\[ V_\ev = \langle \eta_1\eta_2\eta_3, \eta_1, \eta_2, \eta_3 \rangle;
\qquad
   V_\od = \langle \eta_2\eta_3, \eta_3\eta_1, \eta_1\eta_2 \rangle \]
and dim $\fg_0 = (12|12)$.

The elements of $\fg_{-1}$ and $\fg_0$ can be expressed as
vector fields in $\fvect(4|3)$. Choosing
\[
   \eta_1\eta_2\eta_3 \simeq -\partial_y; \quad
    \eta_i \simeq -\partial_{u_i}; \quad
   \frac{\partial \eta_1\eta_2\eta_3}{\partial\eta_i}\simeq -\partial_{\xi_i}.
\]
it is subject to straightforward verification that the elements of
$\fg_0$, expressed as elements of $\fvect (4|3)$ are of the form:
$$
\begin{matrix}
\partial_{\eta_{1}}  \simeq -y\partial_{\xi_{1}}-\xi_{2}
\partial_{u_{3}}+\xi_{3}\partial_{u_{2}} \\
\partial_{\eta_{2}}  \simeq -y\partial_{\xi_{2}}-\xi_{3}
\partial_{u_{1}}+\xi_{1}\partial_{u_{3}} \\
\partial_{\eta_{3}}  \simeq -y\partial_{\xi_{3}}-\xi_{1}
\partial_{u_{2}}+\xi_{2}\partial_{u_{1}}\end{matrix}
\qquad\begin{matrix}
-\eta_{1}\partial_{\eta_{1}}  \simeq u_{1}\partial_{u_{1}}+
\xi_{2}\partial_{\xi_{2}}+\xi_{3}\partial_{\xi_{3}}+
y\partial_{y} \\
-\eta_{2}\partial_{\eta_{2}}  \simeq u_{2}\partial_{u_{2}}+
\xi_{1}\partial_{\xi_{1}}+\xi_{3}\partial_{\xi_{3}}+
y\partial_{y} \\
-\eta_{3}\partial_{\eta_{3}}  \simeq u_{3}\partial_{u_{3}}+
\xi_{1}\partial_{\xi_{1}}+\xi_{2}\partial_{\xi_{2}}+
y\partial_{y}\end{matrix}
$$
$$
\begin{matrix}
\eta_{1}\partial_{\eta_{2}}  \simeq -u_{2}\partial_{u_{1}}+
\xi_{1}\partial_{\xi_{2}} \\
\eta_{2}\partial_{\eta_{3}}  \simeq -u_{3}\partial_{u_{2}}+
\xi_{2}\partial_{\xi_{3}} \\
\eta_{3}\partial_{\eta_{1}}  \simeq -u_{1}\partial_{u_{3}}+
\xi_{3}\partial_{\xi_{1}} \end{matrix}
\qquad\begin{matrix}
\eta_{2}\partial_{\eta_{1}}  \simeq -u_{1}\partial_{u_{2}}+
\xi_{2}\partial_{\xi_{1}} \\
\eta_{3}\partial_{\eta_{2}}  \simeq -u_{2}\partial_{u_{3}}+
\xi_{3}\partial_{\xi_{2}} \\
\eta_{1}\partial_{\eta_{3}} \simeq -u_{3}\partial_{u_{1}}+
\xi_{1}\partial_{\xi_{3}} \end{matrix}
\qquad
\begin{matrix}
\eta_{1}\eta_{2}\eta_{3}\partial_{\eta_{1}}
\simeq -u_{1}\partial_{y} \\
\eta_{1}\eta_{2}\eta_{3}\partial_{\eta_{2}}
\simeq -u_{2}\partial_{y} \\
\eta_{1}\eta_{2}\eta_{3}\partial_{\eta_{3}}
\simeq -u_{3}\partial_{y} \end{matrix}
$$
$$
\begin{matrix}
\eta_{1}\eta_{2}\partial_{\eta_{3}}  \simeq -u_{3}
\partial_{\xi_{3}} \\
\eta_{2}\eta_{3}\partial_{\eta_{1}}  \simeq -u_{1}
\partial_{\xi_{1}} \\
\eta_{3}\eta_{1}\partial_{\eta_{2}}  \simeq -u_{2}
\partial_{\xi_{2}}
\end{matrix}
\quad\ \
\begin{matrix}
\eta_{1}\eta_{2}\partial_{\eta_{1}}
\simeq -u_{1}\partial_{\xi_{3}}-\xi_{2}\partial_{y} \\
\eta_{2}\eta_{3}\partial_{\eta_{2}}
\simeq -u_{2}\partial_{\xi_{1}}-\xi_{3}\partial_{y} \\
\eta_{3}\eta_{1}\partial_{\eta_{3}}
\simeq -u_{3}\partial_{\xi_{2}}-\xi_{1}\partial_{y}
\end{matrix}
\quad\ \
\begin{matrix}
\eta_{1}\eta_{2}\partial_{\eta_{2}}
\simeq -u_{2}\partial_{\xi_{3}}+\xi_{1}\partial_{y} \\
\eta_{2}\eta_{3}\partial_{\eta_{3}}
\simeq -u_{3}\partial_{\xi_{1}}+\xi_{2}\partial_{y} \\
\eta_{3}\eta_{1}\partial_{\eta_{1}}
\simeq -u_{1}\partial_{\xi_{2}}+\xi_{3}\partial_{y}
\end{matrix}
$$

\ssec{1.2} Now we will give a more explicit description of
$\fvect(0|3)_*$. It will turn out that $\fvect(0|3)_*$ is isomorphic
to $\fle(3)$ as Lie superalgebra; however considered as $\Zee$-graded
algebras we have to define a different grading.
The $\Zee$-graded Lie superalgebra $\fle(3;3)$ is
  $\fle(3)$ as Lie superalgebra with $\Zee$-degree of $D$
\[ D=\sum\limits_i f_i\pder{u_{i}} + \sum\limits_j g_j\pder{\xi_{j}}
\]
the $u$-degree of $f_i$ minus 1 (or the $u$-degree of $g_j$), i.e. deg
$\xi_i=0$.

Consider the map $i_1 : \fle(3; 3)\to \fvect(4|3)$ given by
\begin{itemize}
\item[a.)] If $f=f(u)$ then
$$
  i_1(\Le_{f})=\Le_{\sum\pderf{f}{u_i}\xi_j\xi_k-yf}
$$
where $y$ is treated as a parameter and $(i,j, k)\in A_3$ (even
permutations of $\{1,2,3\})$.
\item[b.)] If $f = \sum f_i(u)\xi_i$ then
$$
  i_1(\Le_f)=\Le_f - \varphi(u)\sum \xi_i\partial_{\xi_i}
             +\left(-\varphi(u)y + \Delta
(\varphi(u)\xi_1\xi_2\xi_3)\right)\partial_y
$$
where $\varphi(u)=\Delta(f)$ and $\Delta$ as given in section 0.4.
\item[c.)] If $f=\psi_1(u)\xi_2\xi_3 + \psi_2(u)\xi_3\xi_1 +
                 \psi_3(u)\xi_1\xi_2$ then
$$
   i_1(\Le_f) = -\Delta(f)\partial_y
-\sum\limits_{i=1}^3\psi_i(u)\pder{\xi_{i}}.
$$
\item[d.)] If $f=\psi(u)\xi_1\xi_2\xi_3$ then
$$
   i_1(\Le_f)=-\psi(u)\partial_y.
$$
\end{itemize}
Note that $i_1$ preserves the $\Zee$-degree. We have the following
lemma.
\ssbegin{1.3}{Lemma} The map $i_1$ is an isomorphism of $\Zee$-graded
Lie superalgebras between $\fle(3;3)$ and $\fvect(0|3)_* \subset
\fvect(4|3)$.
\end{Lemma}
\begin{proof} That $i_1$ is an embedding can be verified by direct
  computation. To prove that the image of $i_1$ is in $\fvect(0|3)_*$
  it is enough to show that this is the case on the components
  $\fle(3; 3)_{-1}\oplus\fle(3; 3)_0$, i.e. on functions $f(u, \xi)$
  of degree $\leq 1$ with respect to $u$, as the Cartan prolongation is
  the biggest subalgebra $\fg$ of $\fvect(4|3)$, with given  $\fg_{-1}$
  and $\fg_0$. The proof that $i_1$ is surjective onto
  $\fvect(0|3)_*$ is given in corollary 4.6.
\end{proof}
A generalized version of Lemma 1.3 can be found in \cite{Sh3} and
\cite{LSh}. It states that $\fle(n;n)$ and $\fvect(0|n)_*$ are
isomorphic for all $n \geq 1$.

\section*{\S 2. The construction of \protect $\fcvect (0|3)_*$}

\ssec{2.1} Let us describe a general construction, which leads to
several new simple Lie superalgebras. Let $\fu=\fvect(m|n)$, let
$\fg=(\fu_{-1}, \fg_0)_*$ be a simple Lie subsuperalgebra of $\fu$.
Moreover suppose there exists an element $d\in\fu_0$ that determines an
exterior derivation of $\fg$ and has no kernel on $\fu_+$.  Let us
study the prolong $\tilde\fg=(\fg_{-1}, \fg_0\oplus \Cee d)_*$.

\begin{Lemma} Either $\tilde\fg$ is simple or $\tilde\fg=\fg\oplus
\Cee d$.
\end{Lemma}
\begin{proof} Let $I$ be a nonzero graded ideal of $\tilde\fg$. The
  subsuperspace $(\ad~ \fu_{-1})^{k+1} a$ of $\fu_{-1}$ is nonzero for
  any nonzero homogeneous element $a\in\fu_k$ and $k\geq 0$. Since
  $\fg_{-1}=\fu_{-1}$, the ideal $I$ contains nonzero elements from
  $\fg_{-1}$; by simplicity of $\fg$ the ideal $I$ contains the whole
  $\fg$. If, moreover, $[\fg_{-1}, \tilde \fg_1]=\fg_0$, then by
  definition of the Cartan prolongation $\tilde\fg=\fg\oplus \Cee d$.

  If, instead, $[\fg_{-1}, \tilde \fg_1]=\fg_0\oplus \Cee d$, then
  $d\in I$ and since $[d, \fu_+]=\fu_+$, we derive that $I=\tilde\fg$. In
  other words, $\tilde\fg$ is simple.
\end{proof}

As an example, take $\fg=\fsvect(m|n)$; $\fg_0=\fsl(m|n)$,
$d=1_{m|n}$. Then $(\fg_{-1}, \fg_0\oplus \Cee d)_*=\fvect(m|n)$.

\ssbegin{2.2}{Definition}
The Lie superalgebra $\fcvect(0|3)_*\subset \fvect(4|3)$ is the Cartan
prolongation with $\fcvect(0|3)_{-1}=\fvect(0|3)_{-1}$ and
$\fcvect(0|3)_{0}=\fvect(0|3)_{0}\oplus \Cee d$, with
$$
 d=\sum u_{i}\partial_{u_{i}}+\sum \xi_{i}\partial_{\xi_{i}}+y\partial_{y}.
$$
\end{Definition}
If now
$$
f=\sum_{i=1}^3\xi_{i}\partial_{\xi_{i}}+2y\partial_{y},
$$
then it is clear that $f\in\fvect(0|3)\oplus \Cee d$, but
$f\not\in\fvect(0|3)$.

\ssbegin{2.3}{Theorem}
The Lie superalgebra $\fcvect (0|3)_*$ is  simple.
\end{Theorem}
\begin{proof}
We know that $\fvect (0|3)_* \cong \fle(3;3)$ is simple. According to
Lemma 2.1 it is sufficient to find an element $F \in \fcvect(0|3)_1$,
which is not in $\fvect (0|3)_1$. For $F$ one can take
\[
F=y\xi_{1}\partial_{\xi_{1}}+y\xi_{2}
\partial_{\xi_{2}}+y\xi_{3}\partial_{\xi_{3}}+
y^{2}\partial_{y}-\xi_{1}\xi_{2}\partial_{u_{3}}-
\xi_{3}\xi_{1}\partial_{u_{2}}-\xi_{2}\xi_{3}\partial_{u_{1}}
\]
Indeed, one easily checks that $\partial_{y}F=f$, while
\[
[\partial_{\xi_{i}},F]  = -\partial_{\eta_{i}}\qquad (i=1,2,3),
\]
and moreover $[\partial_{u_{i}}, F]= 0$. This proves the claim.
\end{proof}
Similar constructions are possible for general $n$. For $n=2$ we obtain
$\fcvect(0|2)_* \cong \fvect(2|1)$, while for $n>3$ one can prove that
$\fcvect(0|n)_*$ is not simple. For details, we refer to \cite{Sh3}.

\ssbegin{2.4}{Lemma} A vector field
$$
  D=\sum\limits_{i=1}^3(P_i\partial_{\xi_i} +Q_i\partial_{u_i}) +R\partial_y
$$
in $\fvect(4|3)$ belongs to
$\fcvect(0|3)_*$ if and only if it satisfies the following system of
equations:
$$
\pderf{Q_i}{u_j}+(-1)^{p(D)}\pderf{P_j}{\xi_i}=0 \text{ for any }i\neq
j;\eqno{(2.1)}
$$
$$
\pderf{Q_i}{u_i}+(-1)^{p(D)}\pderf{P_i}{\xi_i}=\frac{1}{2}\left(\sum_{1\leq
j\leq 3} \pderf{Q_j}{u_j}+\pderf{R}{y}\right )\text{ for }i=1, 2,
3;\eqno{(2.2)}
 $$
$$
\pderf{Q_i}{\xi_j}+\pderf{Q_j}{\xi_i}=0 \text{ for any }i, j;\text{ in
particular
}\pderf{Q_i}{\xi_i}=0;\eqno{(2.3)}
$$
$$
\pderf{P_i}{u_j}-\pderf{P_j}{u_i}=
-(-1)^{p(D)}\pderf{R}{\xi_k} \eqno{(2.4)}
$$
for any $k$ and any even permutation $\begin{pmatrix} 1& 2& 3\\ i& j&
k\end{pmatrix}$.
$$
\pderf{Q_i}{y}=0\text{ for }i=1, 2, 3;\eqno{(2.5)}
$$
$$
\pderf{P_k}{y}=
(-1)^{p(D)}\frac 12\big(\pderf{Q_i}{\xi_j}-\pderf{Q_j}{\xi_i}\big) \eqno{(2.6)}
$$
for any $k$ and for any even permutation $\begin{pmatrix} 1& 2& 3\\ i&
j& k\end{pmatrix}$.
\end{Lemma}

\begin{proof} Denote by $\fg=\oplus_{i\geq -1}\fg_i$ the superspace of
  solutions of the system (2.1)--(2.6). Clearly,
  $\fg_{-1}\cong\fvect(4|3)_{-1}$. We directly verify that the images
  of the elements from $\fvect(0|3)\oplus \Cee d$ satisfy
  (2.1)--(2.6).  Actually, we composed the system of equations
  (2.1)--(2.6) by looking at these images.

The isomorphism $\fg_0=\fvect(0|3)\oplus \Cee d$ follows from dimension
considerations.

Set
\begin{align*}
  D_{u_{j}}(D) &=\sum\limits_{i\leq3}
  (\pderf{P_i}{u_j}\pder{\xi_i}+\pderf{Q_i}{u_j}\pder{u_i})+\pderf{R}{u_j}
  \pder{y};\\
  D_{y}(D)&=\sum\limits_{i\leq3}
  (\pderf{P_i}{y}\pder{\xi_i}+\pderf{Q_i}{y}\pder{u_i})+\pderf{R}{y}\pder{y};\\
 \tilde D_{\xi_{j}}(D)&=(-1)^{p(D)} \sum\limits_{i\leq3}
    (\pderf{P_i}{\xi_j}\pder{\xi_i}+\pderf{Q_i}{\xi_j}\pder{u_i})
    + (-1)^{p(D)} \pderf{R}{\xi_j}\pder{y}.
\end{align*}
The operators $D_{u_{j}}$, $D_{y}$ and $\tilde D_{\xi_{j}}$, clearly, {\it
commute} with
the
$\fg_{-1}$-action. Observe: the operators {\it commute}, not {\it
super\/}commute.

Since the operators in the equations (2.1)--(2.6) are linear
combinations of only these operators $D_{u_{j}},D_{y}$ and $\tilde
D_{\xi_{j}}$, the definition of Cartan prolongation itself ensures
isomorphism of $\fg$ with $\fcvect(0|3)_*$.
\end {proof}

\ssbegin{2.5}{Remark} The left hand sides of eqs.
  (2.1)--(2.6) determine coefficients of the 2-form $L_D\omega$, where
  $L_D$ is the Lie derivative and $\omega=\sum_{1\leq i\leq
    3}du_id\xi_i$. It would be interesting to interpret the right-hand
  side of these equations in geometrical terms as well.
\end{Remark}

\ssbegin{2.6}{Remark}
Lemma 2.4 illustrates how $\fcvect(0|3)_*$ can be characterized by a
set of first order, constant coefficient, differential operators. This
is a general fact of Cartan prolongations; one just replaces the
linear constraints on $\fg_0$ by such operators. For example, for
$\fvect(0|3)_*$ we have the equations (2.1)--(2.6) {\it and}
$$
\pderf{R}{y}-\sum\limits_{i=1}^3 \pderf{Q_i}{u_i}=0 \eqno{(2.7)}
$$
Indeed, this equation is satisfied by all elements of $\fvect(0|3)_0$,
see section 1.1, but not by $d$.
\end{Remark}

\section*{\protect\S 3. Solution of differential equations
\protect $(2.1)-(2.6)$}

Set $D^3_\xi = \frac{\partial^3}{\partial\xi_1\partial\xi_2\partial\xi_3}$.

\ssbegin{3.1}{Theorem} Every solution of the system
$(2.1)-(2.6)$ is of the form:
$$
\begin{matrix}
D=\Le_f+yA_f-(-1)^{p(f)}\left(y\Delta (f)+y^2D^3_\xi f\right )\partial_y+\\
A_g-(-1)^{p(g)}\left(\Delta (g)+2yD^3_\xi g\right)\partial_y,
\end{matrix}\eqno{(3.1)}
$$
where $f, g\in\Cee [u, \xi]$ are arbitrary and the operator
$A_f$ is given by the formula:
$$
A_f=\frac{\partial^2f}{\partial\xi_{2}\partial
\xi_{3}}\pder{\xi_{1}}+\frac{\partial^2f}{\partial
\xi_{3}\partial
\xi_{1}}\pder{\xi_{2}}+\frac{\partial^2f}{\partial \xi_{1}\partial
\xi_{2}}\pder{\xi_{3}}.\eqno{(3.2)}
$$
\end{Theorem}

\begin{proof} First, let us find all solutions of system $(2.1)$--$(2.6)$ for
which
$Q_1 = Q_2 = Q_3 =0$. In this case the system takes the form
$$
\pderf{P_j}{\xi_i} =0\; \; \text{for}\; \;   i\not=j \eqno{(2.1')}
$$
$$
(-1)^{p(D)}\pderf{P_i}{\xi_i} =\frac{1}{2}\pderf{R}{y}\; \; \text{for}\; \;  i
=
1, 2, 3 \eqno {(2.2')}
$$
$$
\pderf{P_i}{u_j} - \pderf{P_j}{u_i} = -(-1)^{p(D)}\pderf{R}{\xi_k}\;  \;
\text{for}\; \;  (i, j, k)\in A_3  \eqno {(2.4')}
$$
$$
\pderf{P_k}{y} = 0 \; \; \text{for}\; \; k = 1, 2, 3 \eqno {(2.6')}
$$
{}From $(2.1')$, $(2.2')$ and $(2.6')$ it follows that
$$
P_i = \Psi_i(u_1, u_2, u_3) + \xi_i\varphi(u_1, u_2, u_3),
$$
where $\varphi = \frac{1}{2}(-1)^{p(D)}\pderf{R}{y}$. For brevity we will
write $\Psi_i(u)$ and $\varphi(u)$. Then
$R = (-1)^{p(D)}\cdot 2\varphi(u)y + R_0(u, \xi)$.

Let us expand the 3 equations of type $(2.4')$; their explicit form is:
\begin{align*}
\pderf{R_0}{\xi_1} & = -(-1)^{p(D)}(\pderf{\Psi_2}{u_3} -
\pderf{\Psi_3}{u_2}) + (-1)^{p(D)}(\pderf{\varphi}{u_2}\xi_3 -
\pderf{\varphi}{u_3}\xi_2),\cr
\pderf{R_0}{\xi_2} & = -(-1)^{p(D)}(\pderf{\Psi_3}{u_1} - \pderf{\Psi_1}{u_3})
+
(-1)^{p(D)}(\pderf{\varphi}{u_3}\xi_1 - \pderf{\varphi}{u_1}\xi_3),\cr
\pderf{R_0}{\xi_3} & = -(-1)^{p(D)}(\pderf{\Psi_1}{u_2} - \pderf{\Psi_2}{u_1})
+
(-1)^{p(D)}(\pderf{\varphi}{u_1}\xi_2 - \pderf{\varphi}{u_2}\xi_1).\cr
\end{align*}
The integration of these equations yields
\begin{align*}
R_0 &= (-1)^{p(D)}(\Psi_0(u) - (\pderf{\Psi_2}{u_3} -\pderf{\Psi_3}{u_2})\xi_1
-
\\ & (\pderf{\Psi_3}{u_1} - \pderf{\Psi_1}{u_3})\xi_2 -
(\pderf{\Psi_1}{u_2} - \pderf{\Psi_2}{u_3})\xi_3 -
(\pderf{\varphi}{u_2}\xi_3\xi_1 + \pderf{\varphi}{u_1}\xi_2\xi_3 +
\pderf{\varphi}{u_3}\xi_1\xi_2)) \\
&= (-1)^{p(D)}(\Psi_0(u) +\Delta
(-\Psi_1\xi_2\xi_3 - \Psi_2\xi_3\xi_1-\Psi_3\xi_1\xi_2 -
 \varphi\xi_1\xi_2\xi_3)).
\end{align*}
Therefore, any vector field $D$ with $Q_1 = Q_2 = Q_3 = 0$ satisfying  (2.1) --
(2.6) is of the form
\begin{align*}
D &= \sum^3_{i=1}\Psi_i(u)\partial_{\xi_i} +
\varphi(u)\sum^3_{i=1}\xi_i\partial_{\xi_i}   + (-1)^{p(D)}\\
& \cdot (\Psi_0(u) + \Delta (-\Psi_1\xi_2\xi_3 - \Psi_2\xi_3\xi_1-
\Psi_3\xi_1\xi_2 -
\varphi\xi_1\xi_2\xi_3) +2\varphi(u)y)\partial_y.
\end{align*}
where, as before,
\[ \Delta = \sum\limits_{i=1}^{3} \frac{\partial}{\partial u_i}
\frac{\partial}{\partial \xi_i}.
\]
Set
$$
g(u,\xi) = g_0(u,\xi) -\Psi_1\xi_2\xi_3 - \Psi_2\xi_3\xi_1-
\Psi_3\xi_1\xi_2 - \varphi\xi_1\xi_2\xi_3,
$$
with $\Delta  g_0 = \Psi_0$ and $\deg_\xi (g_0)\le 1$. Then
$$
A_g = \sum_{i=1}^3\Psi_i\partial_{\xi_i} +
\varphi\sum^3_{i=1}\xi_i\partial_{\xi_i};
\quad D^3_\xi g = \varphi\;
\text{and}\;  (-1)^{p(D)} = (-1)^{p(g)+1}
$$
for functions $g$ homogeneous with respect to parity. In the end we get:
$$
\begin{matrix}
D &= A_g + (-1)^{p(D)}(\Delta (g) + 2yD^3_\xi g)\partial_y\\
  &= A_g - (-1)^{p(g)}(\Delta (g) + 2yD^3_\xi g)\partial_y.
\end{matrix}\eqno{(3.3)}
$$
Let us return now to the system (2.1) -- (2.6). Equations
 (2.3), (2.5), (2.6) imply that there exists a function $f(u,\xi)$
(independent of $y$!) such that
$$
Q_i = -(-1)^{p(D)} \pderf{f}{\xi_i}  \; \; \text{for}\; \; i = 1, 2, 3.
$$
Then (2.1) implies that
$$
P_i = \pderf{f}{u_i} + f_i(u, \xi_i, y).
$$
{}From (2.6) it follows that
$$
\pderf{f_i}{y} = \partial_{\xi_j}\partial_{\xi_k}f \; \; \text{for even
permutations}\;
\; (i, j, k)
$$
or
$$
f_i = y(\partial_{\xi_j}\partial_{\xi_k}f) + \tilde P_i(u, \xi_i).
$$
Observe that $\tilde P_i$ satisfy $(2.1')$ and $(2.6')$; hence, in view of
(2.2), $\pderf{\tilde P_i}{\xi_i}$ does not depend on $i$. Therefore, we can
choose $\tilde R$ so that $(\tilde P_i, \tilde R)$ satisfy eqs. $(2.1')$,
$(2.2')$, $(2.4')$, $(2.6')$. Thanks to the linearity of  system    (2.1) --
(2.6)  the vector field $D$ is then of  the form
$$
D = D_f + \tilde D, \eqno {(3.4)}
$$
where  $D_f$ and $\tilde D$ are solutions of  (2.1) -- (2.6) such that
$\tilde D = \sum{\tilde P_i\partial_{\xi_i}} + \tilde R\partial_y$ (i.e.,
$\tilde D$ is of the form (3.3)) and
\begin{align*}
D_f &= \sum (-(-1)^{p(D)}\pderf{f}{\xi_i}\partial_{u_i} +
\pderf{f}{u_i}\partial_{\xi_i}) + \sum y(\partial_{\xi_j}\partial_{\xi_k}f)
\partial_{\xi_i}) + R_f\cdot\partial_y \\
&= \Le_f + yA_f + R_f\partial_y.
\end{align*}
It remains to find $R_f$. Equation (2.2) takes the form
$$
(-1)^{p(D)} yD^3_\xi f=
\frac{1}{2}(-(-1)^{p(D)}(\Delta  f) + \pderf {R_f}{y}).
$$
Hence,
$$
R_f = (-1)^{p(D)}(y^2 D^3_\xi f + y\cdot(\Delta  f) +
R_0(u, \xi)).
$$
Then, we can rewrite (2.4) as
$$
-y\pderf{\Delta  f}{\xi_k} + \pderf{R_0}{\xi_k} =
y\partial_{u_j}\partial_{\xi_j}\partial_{\xi_k}f -
y\partial_{u_i}\partial_{\xi_k}\partial_{\xi_i}f.
$$
Observe that the right hand side of the last equation is
equal to $-y\pderf{\Delta  f}{\xi_k}$. This means that
$\pderf{R_0}{\xi_k} = 0$ or $R_0 = R_0(u)$. Therefore, replacing
$\tilde R$ with $\tilde R + R_0$ we may assume that $R_0 = 0$. Then
$$
D_f = \Le_f + yA_f + (-1)^{p(D)}(y(\Delta  f) +
 y^2D^3_\xi f)\partial_y.
\eqno{(3.5)}
$$
By uniting (3.3) -- (3.5) we get (3.1). \end{proof}

\section*{\protect \S 4 How to generate $\fcvect (0|3)_*$ by pairs of
functions}

We constructed $\fcvect(0|3)_*$ as an extension of
$\fvect(0|3)_*\cong\fle(3; 3)$, see lemma 1.3.  Using the results of
section 3, we obtain another embedding $i_2 : \fle(3) \to
\fvect(0|3)_*$.

\ssbegin{4.1}{Lemma} The map
$$
i_2:\Le_f\to \Le_f+yA_f-(-1)^{p(f)}\left(y\Delta (f)+y^2D^3_\xi
f\right)\partial_y
\eqno{(4.1)}
$$
determines an embedding of
$\fle(3)$ into  $\fcvect(0|3)_*$. This embedding preserves the standard
grading of $\fle(3)$.
\end{Lemma}

\begin{proof} We have to verify the equality
$$
i_2(\Le_{\{f, g\}}) = [i_2(\Le_f), i_2(\Le_g)].
$$
Comparison of coefficients of different powers of $y$ shows that the above
equation is equivalent to the following system:
$$
\Le_{\{f, g\}} = [\Le_f, \Le_g]. \eqno {(4.2)}
$$
$$
A_{\{f, g\}} = [\Le_f, A_g] + [A_f, \Le_g] - (-1)^{p(f)}
(\Delta (f)\cdot A_g + (-1)^{p(f)p(g)}\Delta (g)A_f). \eqno {(4.3)}
$$
$$
[A_f, A_g] = (-1)^{p(f)}\left (D^3_\xi f\cdot A_g +(-1)^{p(f)p(g)}
D^3_\xi gA_f\right ).\eqno {(4.4)}
$$
$$\Delta  (\{f, g\}) = \{\Delta f, g\} - (-1)^{p(f)}\{f, \Delta  g\}.
\eqno{(4.5)}
$$
$$
\begin{matrix}
  D^3_\xi \{f, g\} = \{D^3_\xi f, g\}- (-1)^{p(f)}\{f, D^3_\xi g\}
  -(-1)^{p(f)}(A_f(\Delta g) \\ + (-1)^{p(f)p(g)}A_g(\Delta f))+
  \Delta fD^3_\xi g- D^3_\xi f\Delta g.
\end{matrix}
\eqno{(4.6)}
$$
Equation (4.2) is known, see section 0.3. The equalities (4.3)--(4.6)
are subject to direct verification.
\end{proof}
We found two embeddings $i_1 : \fle(3; 3)\to \fvect(0|3)_*$ and $i_2 :
\fle(3)\to \fcvect(0|3)_*$. Let us denote
$$
   \alpha_g = A_g - (-1)^{p(g)}(\Delta g + 2yD^3_\xi g)\partial_y.
$$
We want to prove that the sum of the images of $i_1$ and $i_2$
cover the whole $\fcvect(0|3)_*$. According to Theorem 3.1, it is
sufficient to represent $\alpha_g$ in the form $\alpha_g= i_1g_1 +
i_2g_2$. For convenience we simply write $f$ instead of $\Le_f$.

\ssbegin{4.2}{Lemma} For $\alpha_g$ we have:
$$
\alpha_g=\left\{\begin{matrix}
0&\text{if}\; \; \deg_\xi g = 0\\
i_1(-(\Delta g)\xi_1\xi_2\xi_3) &\text{if}\; \; \deg_\xi g = 1\\
i_1(g) &\text{if}\; \;\deg_\xi g = 2\\
i_1(-\Delta ^{-1}(D^3_\xi g)) + i_2(\Delta ^{-1}(D^3_\xi g)) &\text{if}\;
\;\deg_\xi g = 3.
\end{matrix}\right .
$$
\end{Lemma}
The {\it right inverse} $\Delta^{-1} $ of $\Delta$ is given in section 0.4.\\
The proof of Lemma 4.2 is a direct calculation.

\ssec{4.3. A wonderful property of $\fsle\degree(3)$} In the standard
grading of $\fg=\fsle\degree(3)$ we have: $\dim
\fg_{-1}=(3|3)$, $\fg_0\cong\fspe(3)$. For the regraded superalgebra
$R\fg=\fsle\degree(3;3)\subset \fle(3;3)$ we have: $\dim
R\fg_{-1}=(3|3)$, $R\fg_0=\fsvect(0|3)\cong\fspe(3)$. For the
definition of $\fspe(3)$ we refer to \cite{K} or \cite{Sh3}. Therefore, for
$\fsle\degree(3)$ and only for it among the $\fsle\degree(n)$, the
regrading $R$ determines a nontrivial automorphism. In terms of generating
functions the regrading is determined by the formulas:
\begin{itemize}
\item[1)] $\deg_\xi(f)=0$: $R(f)=\Delta (f\xi_1\xi_2\xi_3)$;
\item[2)] $\deg_\xi(f)=1$: $R(f)=f$;
\item[3)] $\deg_\xi(f)=2$: $R(f)=D^3_\xi (\Delta^{-1}f)$.
\end{itemize}
Note that $R^2(f) = (-1)^{p(f)+1} f$.
Now we can formulate the following proposition.

\ssbegin{4.4}{Proposition}
The {\it nondirect\/} sum of the images of $i_1$ and $i_2$ covers the
whole $\fcvect(0|3)_*$, i.e.,
\[ i_1(\fle(3;3))+i_2(\fle(3))=(\fcvect(0|3))_{*}. \]
We also have
$$
i_1(\fle(3; 3))\cap i_2(\fle(3))\cong\fsle\degree(3; 3)\cong\fsle\degree(3).
$$
\end{Proposition}
\begin{proof}
  The first part follows from Lemma 4.2. The second part follows by
  direct calculation from solving $i_2(\Le_f)=i_1(\Le_g)$. Note that
  $\Le_f \in \fsle\degree(3)$ iff $\Delta(f)=0$ and $D^3_\xi f=0$, and
  similar for $\Le_g \in \fsle\degree(3;3)$. The equation
  $i_2(\Le_f)=i_1(\Le_g)$ is only solvable if $f \in
  \fsle\degree(3)$ and $g \in \fsle\degree(3;3)$, and in this case we
  obtain $g=(-1)^{p(f)+1}Rf$.
\end{proof}
Therefore, we can identify the space of the Lie superalgebra
$\fcvect(0|3)_*$ with the quotient space of $\fle(3; 3)\oplus \fle(3)$
modulo
\[
   \{(-1)^{p(g)+1}Rg\oplus(-g), g\in\fsle\degree (3)\}.
\]
In other words, we can represent the elements of $\fcvect(0|3)_*$ in
the form of the pairs of functions
$$
   (f, g), \quad\text{where}\quad f, g\in \Pi\Cee[u,\xi]/\Cee\cdot 1
\eqno{(4.7)}
$$
subject to identifications
$$
   (-1)^{p(g)+1} (Rg, 0)\sim(0, g)\quad\text{for any}\quad g\in\fsle\degree
(3).
$$

\ssbegin{4.5}{Corollary}  The map $\varphi$ defined by the formula
$$
\varphi|_{i_1(\fle(3; 3))}=\sign\ i_2i_1^{-1};\quad \quad
\varphi|_{i_2(\fle(3))}=i_1 i_2^{-1}
$$
is an automorphism of $\fcvect(0|3)_*$. Here
 $\sign(D) = (-1)^{p(D)} D$.
\end{Corollary}
The map $\varphi$ may be represented in inner coordinates of $\fvect(4|3)$
as a regrading by setting $\deg y = -1; \deg u_i = 1; \deg \xi_i = 0$.

In the representation (4.7) we have
$$
\varphi(f, g)=(g, (-1)^{p(f)+1} f).
$$
Now we can complete the proof of Lemma 1.3.

\ssbegin{4.6}{Corollary} The embedding $i_1 : \fle(3) \to \fcvect(0|3)_*$ is
a surjection onto $\fvect(0|3)_*$.
\end{Corollary}
\begin{proof}
  By Proposition 4.4 we merely have to prove that $i_2(\Le_f)
  \in \fvect(0|3)_*$ iff $\Delta f = 0$ and $D_\xi^3 f = 0$.
  Applying equation (2.7) to $i_2(\Le_f)$, this follows immediately.
\end{proof}

\section*{\protect \S 5 The bracket in $\fcvect (0|3)_*$}

Now we can determine the bracket in $\fcvect(0|3)_*$ in terms of
representation $(f,g)$ as stated in formula (4.7).

We do this via $\alpha_g$. By Theorem 3.1 any $D\in\fcvect(0|3)_*$ is
of the form $D = i_2(f) + \alpha_g$ for some generating functions $f$
and $g$. To determine the bracket $[i_2(f),i_1(h)]$, we
\begin{enumerate}
\item  Compute the brackets  $[i_2f, \alpha_g]$ for any
$f, g \in \Cee[u, \xi]/\Cee\cdot 1;$
\item Represent  $i_1(h)$ in the form
$$
i_1(h) = i_2 a(h) + \alpha_{b(h)} \; \text{for any}\;
h\in\Cee[u, \xi]/\Cee \cdot 1;\eqno{(5.1)}
$$
\end{enumerate}
In Lemma 4.2 we expressed $\alpha_g$ in $i_1$ and $i_2$.

\begin{rem*}{Remark} The functions $a(h)$ and $b(h)$ above
  are not uniquely defined. Any representation will do.
\end{rem*}

\ssbegin{5.1}{Lemma} For any functions
$f, g\in \Cee[u, \xi]/\Cee\cdot 1$ the bracket $[i_2f, \alpha_g]$ is of the
form
$$
[i_2f, \alpha_g] = i_2F + \alpha_G,\eqno{(5.2)}
$$
where
\[
F = f\cdot D^3_\xi g - (-1)^{(p(f) +1)(p(g) + 1)} A_gf \quad
\text{ and }\quad G = -f\Delta  g
\]
\end{Lemma}
\begin{proof} Direct calculation gives that
\begin{align*}
[i_2f, \alpha_g] &= [\Le_f, A_g] +
(-1)^{p(f)p(g) + p(f) +1}\Delta  g\cdot A_f
\\
+ &\ y\left(  [A_f, A_g] + (-1)^{p(f)p(g) +p(f) + 1}\cdot 2\cdot
D^3_\xi g\cdot A_f\right) \\
+ & \ (-1)^{p(g) + 1} \left(\{f, \Delta  g\} + (-1)^{p(f)}\Delta
    f\cdot\Delta  g  \right) \partial_y \\
+ &\ \left(  (-1)^{p(g) + 1}A_f(\Delta  g)  + (-1)^{p(f)p(g) + p(g) +
  1}A_g(\Delta  f) \right.\\
& +\left.
2\cdot(-1)^{p(g)+1}\{f, D^3_\xi g\} +2 \cdot(-1)^{p(f)+p(g)+1}D_\xi^3 f\cdot
\Delta  g \right) y\partial_y \\
+ & (-1)^{p(f)+p(g)+1}2\cdot D^3_\xi f\cdot D^3_\xi g\cdot y^2\partial_y.
\end{align*}
In order to find the functions
$F$ and $G$, it suffices to observe that the coefficient of
$\partial_y$, non-divisible by $y$, should be equal to
$(-1)^{p(G)+1}\Delta  G$. This implies the equations:
$$
(-1)^{p(G)+1}\Delta  G = (-1)^{p(g)+1}\left( \{f, \Delta  g\} + (-1)^{p(f)}
\Delta  f\cdot\Delta  g \right)
$$
or
$$
(-1)^{p(G)+1}\Delta  G = (-1)^{p(f) + p(g) +1}\Delta (f\cdot\Delta  g).
$$
Here $p(G) = p(f\cdot\Delta  g) = p(f) + p(g) + 1$.
Hence, $\Delta  G = \Delta (-f\Delta  g)$. Since
$G$ is defined up to elements from $\fsle\degree (3)$, we can take  $G =
-f\Delta  g$.

The function $F$ to be found is determined from the equation
$$
i_2F = [i_2f, \alpha_g] - \alpha_G. \eqno {(5.3)}
$$
By comparing the coefficients of $y\partial_y$ in the left and right hand sides
of (5.3) we get
\begin{align*}
(-1)^{p(F)+1}\Delta  F &= (-1)^{p(g)+1}A_f(\Delta  g) + (-1)^{p(f)p(g)
+p(g)+1} A_g(\Delta  f)\\
&+ 2(-1)^{p(g)+1}\{f, D^3_\xi g\} +
(-1)^{p(f)+p(g)+1}  2\cdot D^3_\xi f\cdot\Delta  g\\
&- 2\cdot(-1)^{p(f)+p(g)} D^3_\xi(-f\Delta  g).
\end{align*}
Observe that
\begin{align*}
D^3_\xi(f\Delta  g) &= (D^3_\xi f)\Delta  g + (-1)^{p(f)}A_f(\Delta
g) + \sum_{i=1}^{3}\pderf{f}{\xi_i}\pder{u_i}(D^3_\xi g)\\
&= (D^3_\xi f)\cdot\Delta  g + (-1)^{p(f)} A_f(\Delta  g) +(-1)^{p(f)}\{f,
D^3_\xi g\}.
\end{align*}
Then
$$
(-1)^{p(F)+1}(\Delta  F) = (-1)^{p(g)} A_f(\Delta  g) +
(-1)^{p(f)p(g) + p(g) + 1} A_g(\Delta  f).
$$
By comparing parities we derive that
$$
p(F) + 1 = p(A_f(\Delta  g)) = p(f) + 1 + p(g) + 1 = p(f) + p(g).
$$
It follows that
$$
\Delta  F = (-1)^{p(f)}A_f(\Delta  g) + (-1)^{p(f)p(g)+p(f)+1}
A_g(\Delta  f).
$$
Let us transform the right hand side of the equality obtained. The sums
over $i,j,k$ are over $(i, j, k)\in A_3$:
\begin{align*}
   & (-1)^{p(f)}A_f(\Delta  g) + (-1)^{p(f)p(g)+p(f)+1} A_g(\Delta  f)\\
= & \sum (-1)^{p(f)}\partial_{\xi_j} \partial_{\xi_k}f\cdot\partial_{\xi_i}
     (\sum^3_{s=1}\partial_{u_s}\partial_{\xi_s}g)\\
& + (-1)^{p(f)p(g)+p(f)+1} \sum
\partial_{\xi_j}\partial_{\xi_k}g\partial_{\xi_i}
    (\sum^3_{s=1}\partial_{u_s} \partial_{\xi_s}f)\\
= & (-1)^{p(f)p(g)+p(f)}\cdot \sum
     ((\partial_{u_j}\partial_{\xi_i}\partial_{\xi_j} g +
      \partial_{u_k}\partial_{\xi_i}\partial_{\xi_k}g)\cdot
      \partial_{\xi_j}\partial_{\xi_k}f) \\
& - (-1)^{p(f)p(g)+p(f)}\cdot \sum
      (\partial_{\xi_j}\partial_{\xi_k}g
       (\partial_{u_j}\partial_{\xi_i}\partial_{\xi_j}f +
       \partial_{u_k}\partial_{\xi_i}\partial_{\xi_k}f))\\
= &
(-1)^{p(f)p(g)+p(f)}\sum \partial_{u_k} (\partial_{\xi_i}\partial_{\xi_k}g\cdot
\partial_{\xi_j}\partial_{\xi_k}f +
\partial_{\xi_j}\partial_{\xi_k}g\cdot
\partial_{\xi_k}\partial_{\xi_i}f)\\
= & (-1)^{p(f)p(g)+p(f)+
p(g)}\sum^3_{k=1}(\partial_{u_k}\partial_{\xi_k}(A_gf)-
    \partial_{u_k}{D^3_\xi}g\cdot\partial_{\xi_k}f)) \\
= & -(-1)^{(p(f)+1)(p(g)+1)}\Delta (A_gf) + (-1)^{p(f)p(g)+p(f)}
\Delta (D^3_\xi g \cdot f) \\
= & \Delta (f\cdot D^3_\xi g) - (-1)^{(p(f)+1)(p(g)+1)}\Delta (A_gf).
\end{align*}
Then
$$
F = f\cdot D^3_\xi g - (-1)^{(p(f)+1)(p(g)+1)}A_gf +
F_0,\quad\text{where}\quad
\Delta  F_0 = 0.
$$

We have shown how to {\it find} functions $F$ and $G$. To prove Lemma 5.1
it only remains to compare the elements of the same degree in $y$ in the
right-hand and the left-hand side, i.e., to verify the following three
equalities:
\begin{align*}
(-1)^{p(F)+1}D^3_\xi F =& 2(-1)^{p(f)+p(g)+1} D^3_\xi
f\cdot D^3_\xi g \\
\Le_F + A_G =& [\Le_f, A_g] + (-1)^{p(f)p(g)+p(f)+1}\Delta  g\cdot A_f \\
A_F =& [A_f, A_g] + 2\cdot(-1)^{p(f)p(g)+p(f)+1}D^3_\xi g\cdot A_f
\end{align*}
The verification is a direct one.
\end{proof}

\ssbegin{5.3}{Lemma} The representation of $i_1h$ in the form $(5.1)$ is
as follows:
$$
i_1h=\left\{\begin{matrix}
i_2(\Delta (h\xi_1\xi_2\xi_3)) &\text{if}\; \;  \deg_\xi h = 0,\\
i_2h + \alpha_{(\Delta  h)\xi_1\xi_2\xi_3} &\text{if}\; \; \deg_\xi
h = 1,\\
\alpha_h  &\text{if}\; \; \deg_\xi h = 2,\\
\alpha_{\Delta ^{-1}(D^3_\xi h)}   &\text{if}\; \;   \deg_\xi h
=3.
\end{matrix}\right .
$$
\end{Lemma}

\begin{proof} It suffices to compare the definition of $\alpha_g$
  with the definitions of $i_1$ and $i_2$. If $\deg_\xi h = 0$
  use the equalities $\sum\pderf{f}{u_i}\xi_j\xi_k = \Delta
  (f\xi_1\xi_2\xi_3)$ and $A_{\Delta (f\xi_1\xi_2\xi_3)} = \Le_f$. In
  the remaining cases the verification is not difficult.
\end{proof}

Making use of the Lemmas 5.1, Lemma 5.2 and Lemma 4.2 we can compute
the whole multiplication table of $[i_2f, i_1h]$:
\begin{itemize}
\item $\deg_\xi h = 0$. Then
\[i_1h = i_2(\Delta (h\xi_1\xi_2\xi_3)) \text{ and }
  [i_2f, i_1h] = i_2\{f, \Delta (h\xi_1\xi_2\xi_3)\}.
\]
We also have
$$
\{f, \Delta (h\xi_1\xi_2\xi_3)\} = \left\{\begin{matrix}
0 &\text{if}\; \deg_\xi f=3\\
-\{\Delta  f, h\}\xi_1\xi_2\xi_3  &\text{if}\;  \deg_\xi f = 2.
\end{matrix}\right .
$$
\item $\deg_\xi h = 1$. Then
$$
\begin{matrix}
[i_2f, i_1h] = [i_2f, i_2h + \alpha_{(\Delta  h)\xi_1\xi_2\xi_3}] =\\
i_2\{f, h\} -  i_2(f\Delta  h) + i_2(\Delta  h\cdot
\sum\xi_i\partial_{\xi_i}f) +
\alpha_{-f\cdot\Delta ((\Delta  h)\xi_1\xi_2\xi_3)}.
\end{matrix}
$$
\item $\deg_\xi h = 2$. Then
$$
\begin{matrix}
[i_2f, i_1h] = [i_2f, \alpha_h] =
 (-1)^{p(f)} i_2(A_h f) - \alpha_{(f\Delta  h)}= \\
\left\{\begin{matrix}
i_1(\{f, \Delta
h\}\xi_1\xi_2\xi_3) &\text{if}\;  \deg_\xi f = 0\\
i_1(\Delta (fh) - f\Delta
h) &\text{if}\;   \deg_\xi f = 1\\
i_2(A_hf) -i_2(\Delta ^{-1} D^3_\xi(f\Delta
h)) + i_1(\Delta ^{-1}D^3_\xi(f\Delta  h))   &\text{if}\;  \deg_\xi f =
2\\
-i_2(h D^3_\xi f)   &\text{if}\;  \deg_\xi f = 3.
\end{matrix}\right .
\end{matrix}
$$
\item $\deg_\xi h = 3$. Then
$$
\begin{matrix}[i_2f, i_1h] = [i_2f, \alpha_{\Delta ^{-1}(D^3_\xi h)}] =
-\alpha_{f\cdot D^{3}_{\xi}h} = \\
\left\{\begin{matrix}0 &\text{if}\; \deg_\xi f = 0\\
i_1(-\Delta (f\cdot
D^3_\xi h)\xi_1\xi_2\xi_3) = i_1(-f\Delta  h -\Delta  f\cdot h)  &\text{if}\;
\deg_\xi f = 1\\
i_1(-f\cdot D^3_\xi g) &\text{if}\; \deg_\xi f = 2\\
i_1(\Delta ^{-1}(D^3_\xi f\cdot D^3_\xi g))
-i_2(\Delta ^{-1}(D^3_\xi f\cdot D^3_\xi g)) &\text{if}\; \deg_\xi f
= 3.\end{matrix}\right .
\end{matrix}
$$
\end{itemize}
The final result is represented in the following tables.

\section*{The brackets \protect$[i_2f, i_1h]$}
$$
\begin{tabular}{|c||c|c|}
\hline
$\deg_\xi(f)$&$\deg_\xi(h)=0$&$\deg_\xi(h)=1$\\
\hline\hline
$0$ & $i_2(\{f, \Delta(h\xi_1\xi_2\xi_3)\})$ &
$-i_1(\{\Delta(f\xi_1\xi_2\xi_3),h\})$\\
\hline
$1$&$i_2(\{f, \Delta(h\xi_1\xi_2\xi_3)\})$&$i_1(\Delta^{-1}\{f,\Delta h\})+$ \\
   &   &$i_2(\{f, h\}-\Delta^{-1}\{f,\Delta h\})$  \\
\hline
$2$ & $-i_2(\{ \Delta f,h\}\xi_1\xi_2\xi_3)$ & $i_2(\Delta(fh)-\Delta(f)h)$\\
\hline
$3$ & $0$ & $i_2(f\Delta(h)+\Delta(f)h)$\\
\hline
\end{tabular}
$$

$$
\begin{tabular}{|c||c|c|}
\hline
$\deg_\xi(f)$ & $\deg_\xi(h)=2$ & $\deg_\xi(h)=3$\\
\hline\hline
$0$ & $i_1(\{f, \Delta h\}\xi_1\xi_2\xi_3)$ & $0$\\
\hline
$1$ & $-i_1(\Delta(fh)+f\Delta h)$ & $i_1(-f\Delta(h)-\Delta(f)h)$\\
\hline
$2$  & $i_1(\Delta^{-1}D^3_\xi(f\Delta h))+$       & $i_1(-fD^3_\xi h)$ \\
     & $i_2(A_hf -\Delta^{-1}D^3_\xi(f\Delta h))$  &                     \\
\hline
$3$ & $i_2(-hD^3_\xi f)$ & $i_1(\Delta^{-1}(D^3_\xi f\cdot D^3_\xi h))-$ \\
    &                    & $i_2(\Delta^{-1}(D^3_\xi f\cdot D^3_\xi h))$  \\
\hline
\end{tabular}
$$

\end{document}